\documentclass[12pt,letter]{article}
\usepackage[margin=1in]{geometry}
\usepackage{amsmath, longtable,graphics, amssymb, float,epsfig,shapepar, amsbsy, amsmath, amsthm, amssymb, color, amsfonts,natbib,indentfirst,graphicx, enumerate,biometrics}
\usepackage{setspace,fancyhdr}
\usepackage{graphicx}
\usepackage{subfigure}
\usepackage{hyperref}
\hypersetup{
    bookmarks=false,         % show bookmarks bar?
    unicode=false,          % non-Latin characters in Acrobat s bookmarks
    pdftoolbar=true,        % show Acrobats toolbar?
    pdfmenubar=true,        % show Acrobats menu?
    pdffitwindow=false,     % window fit to page when opened
    pdfstartview={FitB},    % fits the width of the page to the window
    pdftitle={mpcr},    % title
    pdfauthor={Zhenke Wu},     % author
    pdfsubject={Subject},   % subject of the document
    pdfcreator={Creator},   % creator of the document
    pdfproducer={Producer}, % producer of the document
    pdfkeywords={keywords}, % list of keywords
    pdfnewwindow=true,      % links in new window
    colorlinks=true,%false,       % false: boxed links; true: colored links
    pdfborder={0,0,0},
    linkcolor=blue,%red,          % color of internal links
    citecolor=blue,        % color of links to bibliography
    filecolor=magenta,      % color of file links
    urlcolor=blue,%blue,%cyan,           % color of external links
    pdfpagemode=None
}
\usepackage{float}
\usepackage{multirow,bigdelim}
\usepackage[bottom]{footmisc}

\theoremstyle{plain} 

\theoremstyle{plain}

\theoremstyle{plain} 
\theoremstyle{plain} 
\theoremstyle{plain} 

\theoremstyle{plain} 

\theoremstyle{plain} 

\newcommand{\pr}{\ensuremath{\mbox{pr}}}

\newcommand{\longpage}{\enlargethispage{\baselineskip}}

\newcommand{\effect}{\ensuremath{\mbox{\footnotesize\sf effect } }}
\newcommand{\calibr}{\ensuremath{\mbox{\footnotesize\sf calibr } }}
\newcommand{\crude}{\ensuremath{\mbox{\footnotesize\sf crude } }}

\newcommand{\mucalibr}{\ensuremath{\mu^{\calibr}}}

\renewcommand{\thetable}{\arabic{table}}

\theoremstyle{plain} 
\theoremstyle{plain} 
\theoremstyle{plain}

\theoremstyle{plain}

\renewcommand{\thesection}{\arabic{section}.}
\renewcommand{\thesubsection}{\arabic{section}\mbox{$.$}\arabic{subsection}}
\def\theequation {\arabic{equation}}

\fancypagestyle{title}{%
  \setlength{\headheight}{22pt}%
  \fancyhf{}% No header/footer
  \fancyfoot[C]{\thepage}% Page number in Centre of footer
  \fancyhead[L]{\href{http://onlinelibrary.wiley.com/doi/10.1111/biom.12214/abstract?deniedAccessCustomisedMessage=&userIsAuthenticated=false}{\footnotesize \textit{Biometrics}, doi: 10.1111/biom.12214}}
  \fancyhead[R]{}
}%

\title{\bf \Large{\vspace*{-1.3cm Estimation of Treatment Effects in \\ Matched-Pair Cluster Randomized Trials \\ by Calibrating Covariate Imbalance Between Clusters}}}

\footnotetext[1]{Department of Biostatistics, Johns Hopkins University, Baltimore, MD 21205, U.S.A}
\footnotetext[2]{U.S. Census Bureau, Suitland, Maryland 20746, U.S.A.}

\author{\normalsize Zhenke Wu\(^{1,} \footnote{E-mail:~zhwu@jhu.edu}\), Constantine E. Frangakis\(^{1,} \footnote{E-mail:~cfrangak@jhsph.edu}\), Thomas A. Louis\(^{1,2,} \footnote{E-mail:~tlouis@jhu.edu}\), Daniel O. Scharfstein\(^{1,} \footnote{E-mail:~dscharf@jhsph.edu}\)}
\date{May 26th, 2014}

\bibpunct{(}{)}{;}{a}{,}{,}

\usepackage{fancyhdr}
\begin{document}

\maketitle
\thispagestyle{title}

\lhead{\href{http://onlinelibrary.wiley.com/doi/10.1111/biom.12214/abstract?deniedAccessCustomisedMessage=&userIsAuthenticated=false}{\footnotesize \textit{Biometrics}, doi: 10.1111/biom.12214}}
\rhead{}

\noindent {\sc Summary.} {We address estimation of intervention effects in experimental designs in which (a) interventions are  assigned at the cluster level; (b) clusters are selected to form pairs, matched on observed characteristics; and (c) intervention is assigned to one cluster at random within each pair. One goal of policy interest is to estimate the average outcome if all clusters in all pairs are assigned control versus if all clusters in all pairs are assigned to intervention. In such designs, inference that ignores individual level covariates can be imprecise because cluster-level assignment can leave substantial imbalance in the covariate distribution between experimental arms within each pair. However, most existing methods that adjust for covariates have estimands that are not of policy interest. We propose a methodology that explicitly balances the observed covariates among clusters in a pair, and retains the original estimand of interest. We demonstrate our approach through the evaluation of the Guided Care program.
}

\renewcommand{\thefootnote}{\arabic{footnote}}
\setcounter{footnote}{0}
\vspace*{.3cm}


\maketitle

\newpage
\section{Introduction}
Experimental designs often have the following three features: interventions are assigned at the cluster level; clusters are selected to form pairs, matched on observed covariates; and interventions are assigned to one cluster at random within each pair. One goal of policy interest is to estimate the average outcome if all clusters in all pairs are assigned control versus if all clusters in all pairs are assigned to intervention. The effect of such a policy is easy to understand, because its definition or estimation does not have to depend on models. Such designs are useful when individual-level randomization is not feasible due to practical constraints, and when cluster assignment also reflects how the assignment would scale in practice.

The Guided Care program is a recent example of such a study \citep{Boult}. The study's goal was to assess the effect of Guided Care versus a control condition on functional health and other patient outcomes among clinical practices serving chronically ill older adults. In Guided Care, a trained nurse works closely with patients and their physicians to provide coordinated care.  The control group does not have access to such a nurse. To assess the effect of the Guided Care intervention, the study recruited 14 clinical practices and matched them in 7 pairs using clinical practice and patient characteristics, and within each pair assigned randomly one clinical practice to Guided Care and the other to control.

A problem with cluster-level assignment is that it can leave substantial imbalances in the covariates within pairs. However, existing methods to estimate effects in such designs rarely use covariates in order to adjust for these imbalances. As a consequence, such methods, including nonparametric as well as hierarchical (meta-analysis) approaches, although useful in other ways \citep{Imai09}, can leave large uncertainty in the results. Methods that do use covariates usually estimate effects conditionally on covariates and cluster-specific random effects \citep{Thompson1997,Feng2001,Hill2009}. With such methods, the estimands are no longer of policy interest and lack meaning when the modelling assumptions are misspecified.

We propose an approach that explicitly balances the observed covariates between clusters in a pair and still estimates causal effects of policy interest. In Section 2, we formulate the matched-pair cluster randomized design through potential outcomes. We then characterize in Section 3 the existing approaches to causal effects estimation and their complications. In Section 4, we propose a covariate-calibration approach and develop inferences with and without the need for assumptions for a hierarchical second level. Throughout these sections, the arguments are demonstrated through the evaluation of the recent Guided Care program. Section 5 concludes with discussion.

\section{The goal and design using potential outcomes}
Consider a design that operates in pairs $p = 1, \dots, n$ of clusters. In each pair \(p\), the design recruits two clusters (e.g., clinical practices) indexed by  $i = 1, 2$, matched on qualitative and quantitative characteristics, such as percentage of patients with private insurance, and where each clinical practice serves a community, say with a large number of \(N_{p,i}\) patients. The design then assigns to each clinic one of two treatments, namely control ($t = 1$) or intervention ($t = 2$). If clinical practice $i$ of pair $p$ is assigned treatment $t$, then potential outcomes $Y_{p,i,k}(t)$  \citep{Rubin1974, Rubin1978} are to be measured on a random sample of $k = 1, \dots, n_{p,i}$ patients from the \(N_{p,i}\) patients served in that clinical practice. We label $F_{p,i}(y; t), \mu_{p,i}(t)$, and $\sigma^2_{p,i}(t)$ the distribution (at value \(y\)), mean and variance of the potential outcome $Y_{p,i,k}(t)$ within clinical practice \(i\) of pair \(p\). The average outcomes in pair $p$ are
%One goal of interest is in assessing if intervention causes any difference in average outcomes in a practice \(p\), namely assessing deviations from the null hypothesis that
%\begin{equation}
%H_0: \mu_{p}(1)= \mu_{p}(2), \mbox{ for all \(p\), where  }\mu_{p}(t):=\mu_{p,i=1}(t)\pi_{p,i=1}+\mu_{p,i=2}(t)\pi_{p,i=2}, \label{null}
%\end{equation}
%and where $\pi_{p,i=1}$ is the fraction of patients served by clinic $i=1$, i.e. $N_{p,i=1}/(N_{p,i=1}+N_{p,i=2})$, and similarly for $\pi_{p,i=2}$.
\begin{equation}
\mu_{p}(t):=\mu_{p,i=1}(t)\pi_{p,i=1}+\mu_{p,i=2}(t)\pi_{p,i=2}, \label{null}
\end{equation}
where \(``:="\) means ``define", $\pi_{p,i=1}$ is the fraction of patients served by clinic $i=1$, i.e. $N_{p,i=1}/(N_{p,i=1}+N_{p,i=2})$, and similarly for $\pi_{p,i=2}$.
%Another, more
One goal of policy interest is to estimate the average outcome if all clinical practices in all pairs are assigned control versus if all clinical practices in all pairs are assigned intervention. In terms of  the model, the goal is to estimate a contrast between
\begin{align}
\mu(1):=& E\{\mu_p(t=1)\}\mbox{ and  }\mu(2):=E\{\mu_p(t=2)\},\nonumber\\[-.4cm]
\label{estimands}\\[-.4cm]
&\mbox{ for example } \delta^{\effect}:=\mu(1)-\mu(2),\nonumber
\end{align}
which is the average outcome if all clusters had been assigned treatment 1 versus if all clusters had been assigned treatment 2. Here, the expectations are taken over a larger population \(P\) of pairs from which $p = 1, \dots, n$ can be considered a random sample. Alternative estimands (e.g. conditionally on the sample of pairs, \cite{Imai09}) can be considered, although this does not change the main issues discussed here.
%A related goal is to assess deviations from nulcan be performed. We describe detailed forms of null hypotheses when needed in specific settings.}

Within each pair, the design assigns at random the intervention to one clinical practice and the control to the other, independently across pairs. Because in this design the original ordering $i$ is arbitrary, and in order to ease comparisons with the existing meta-analytic approach (e.g. \cite{Thompson1997}), for each pair $p$ we relabel by $c= 1$ the clinical practice that \textit{is} assigned control, and by $c = 2$ the clinical practice that \textit{is} assigned intervention. The quantities $Y_{p,c,k}(t)$, $F_{p,c}(y;t)$, $\mu_{p,c}(t)$ and $\sigma^2_{p,c}(t)$ are then redefined based on this relabeling and the above definitions. %For each such patient, covariates \(X_{p,i,k}\) have been measured before the assignment to treatment.
  Then, the paired cluster randomized design implies the following:
\vspace{.3cm}

%\noindent {\sc Condition 1. }{\em The distribution of potential outcomes  under treatments 1 and 2 conditionally on covariates,  and the distribution of covariates, in clinic \(c\), namely the vector of functions \(\big [ F_{p,c}(\cdot\mid\cdot,t=1), F_{p,c}(\cdot\mid\cdot,t=2); G_{p,c}(\cdot)), \big ]\) is exchangeable (in distribution over practices) between clinics \(c=1\) and \(c=2\). }

\noindent {\sc Condition 1. }{\em The potential outcomes under treatments 1 and 2 in clinical practice \(c\), and the number of patients served by clinical practice \(c\) are exchangeable (in distribution over pairs) between clinical practices \(c=1\) and \(c=2\), i.e.,

\begin{figure}[H]
\begin{center}
\includegraphics[width=.7\textwidth]{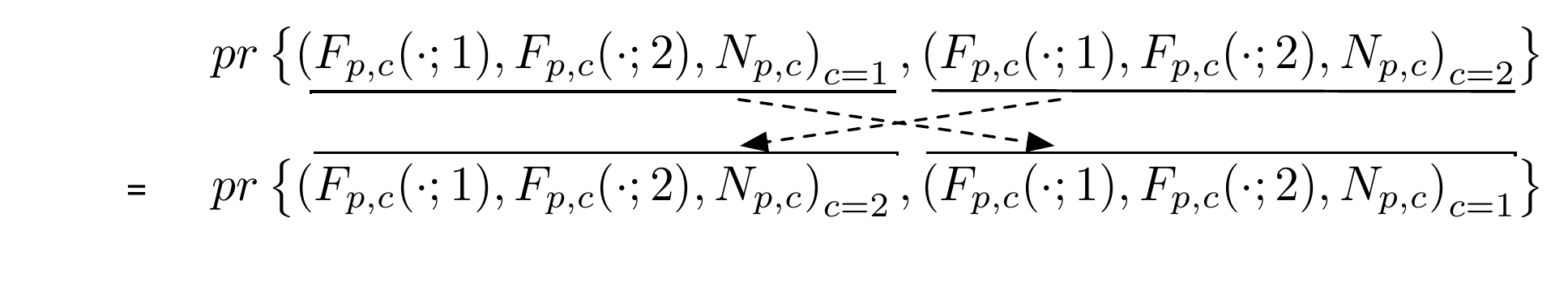}
\end{center}
\label{fig:exch}
\end{figure}
%\vspace*{-1.5cm}
\noindent where the arrows connect equal entries in arguments, and distribution \(\pr\) is over pairs \(p\) in the larger population \(P\) of pairs. }
\vspace*{.3cm}

\begin{figure}[H]
\begin{center}
\hspace*{-.1cm}\includegraphics[width=0.7\textwidth]{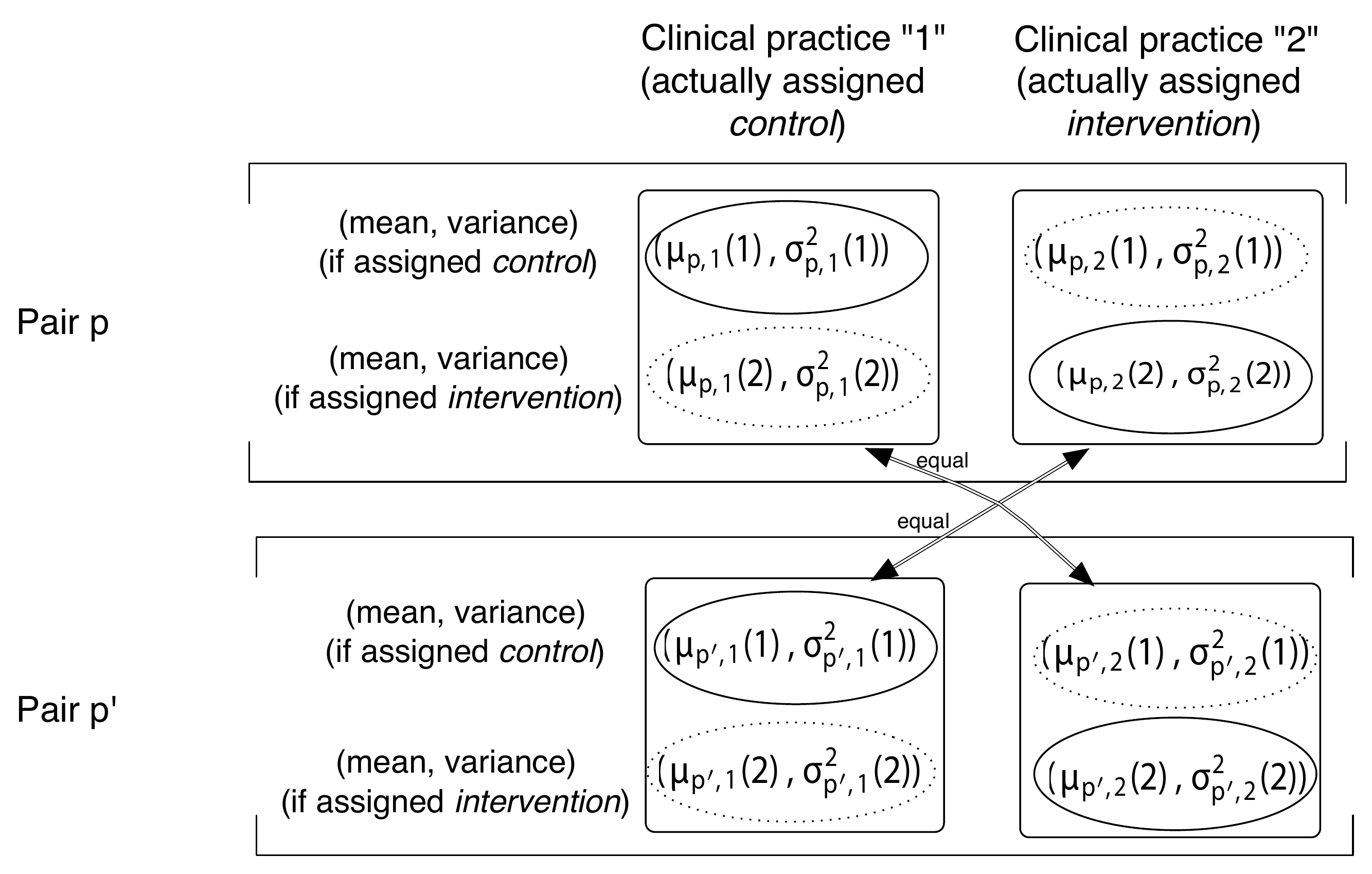}
\end{center}
\vspace{-0.1in}
\caption{The underlying structure of the paired-cluster randomized design. The top part (observed pair \(p\)) and bottom part (observed pair \(p'\)) are the {\em two} possible ways in which a {\em single} pair can be manifested in the design. Observed pair \(p\) has two clinical practices (represented by the two squares). For each clinical practice, the first row shows the mean and variance of patient outcomes if the clinical practice is assigned control and the second row shows the mean and variance if assigned intervention.  The clinical practice {\em actually} assigned control is indicated by its placement in column ``1" , and the clinical practice {\em actually} assigned intervention is in column ``2". The solid (nonsolid) ellipsoids show the means and variances that can (cannot) be estimated directly. Observed pair \(p'\) shows how the same pair would be manifested in the design if the assignment of treatment to clinical practices were in reverse (a line with arrows connects the same clinical practice in these two different assignments). Condition 1 means that each of the two manifestations, \(p\) and \(p'\) has the same probability.}
\label{fig:design}
\end{figure}

Condition 1 implies, for example, that the joint distribution of the ``means and variances under exposure to intervention" is the same for the clinical practices that are actually assigned the intervention (clinical practices ``2") as it is for the clinical practices that are actually assigned the control (clinical practices ``1").  Figure \ref{fig:design} illustrates the structure of pairs, clinical practices, and assigned treatments in this paired cluster randomized design, along with means and variances of potential outcome distributions.

[Figure \ref{fig:design} here.]
%By definition, $Y_{p,c,k}(t)$ will be observed when $t = c$. As in \cite{Thompson1997}, such outcomes will be simply denoted as $Y_{p,c,k}$. Those observed pair of outcomes are shown by connected solid rectangles; those unobserved are shown by connected dashed rectangles.

%\section{Existing methods}
Here we connect the observed data and existing methods to the above framework of potential outcomes, because this helps understand the meaning of the assumptions, explicit or implicit, required by the existing methods.

%\subsection{Complications when ignoring covariates in the analysis. }
In order to estimate an effect such as \(\delta^{\effect}\) of (\ref{estimands}), consider first a particular pair $p$: we can directly estimate the average potential outcome under control for the clinical practice assigned to the control, namely \(\mu_{p,c=1}(t=1)\); and the average potential outcome under intervention for the clinical practice assigned to the intervention, namely \(\mu_{p,c=2}(t=2)\). Specifically, for the control clinical practice ($c=1$) of pair \(p\), let $\hat{\mu}_{p,c=1}(t=1):= \frac{1}{n_{p,c=1}}\sum_{k=1}^{n_{p,c=1}}Y_{p,c=1,k}(t=1)$ denote the average of the observed outcomes, i.e., the potential outcomes under \(t=1\); and for the intervention clinical practice ($c=2$) of pair \(p\), let $\hat{\mu}_{p,c=2}(t=2):= \frac{1}{n_{p,c=2}}\sum_{k=1}^{n_{p,c=2}}Y_{p,c=2,k}(t=2)$ denote the average of the observed outcomes, i.e., the potential outcomes under \(t=2\). Then, letting \(\hat{\delta}^{\crude}_p=\hat{\mu}_{p,1}(1)-\hat{\mu}_{p,2}(2)\), and conditionally on pairs \(p\) whose clinical practices have particular values of \((\delta^{\crude}_p,v_p^{\crude})\), we have that
\begin{align}
&\hspace*{1cm}\pr ( \hat{\delta}^{\crude}_p \mid \delta^{\crude}_p, v_p^{\crude})\thinspace \dot = \thinspace Normal(\delta^{\crude}_p, v_p^{\crude}), \quad\mbox{where} \nonumber\\[-.3cm]
&\label{firstlevel}\\[-.3cm]
&\delta^{\crude}_{p} : = \mu_{p,1}(1)-\mu_{p,2}(2) \mbox{ and }  v_p^{\crude}=\frac{\sigma^2_{p,1}(1)}{n_{p,1}}+\frac{\sigma^2_{p,1}(2)}{n_{p,2}}.\nonumber
%\delta_{j}^{True} &: = & \frac{\mu_{j,2}(2)-\mu_{j,2}(1)+\mu_{j,1}(2)-\mu_{j,1}(1)}{2},\\
%\delta & :=  & E\left[\delta_{j}^{True} \right].
\end{align}
Here, ``$\dot=$" means ``approximately", the notation \(\pr(A_p \mid B_p)\)  and \(E(A_p \mid B_p)\) means the distribution and expectation, respectively, of characteristic \(A_p\) among pairs in the larger population \(P\) that have characteristic \(B_p\) (if \(B_p\) is empty, the distribution and expectation are over all pairs).
\vspace*{.3cm}

\noindent {\em Remark 1. }In a pair, the directly estimable (crude) contrast \(\delta^{\crude}_p\) is not a causal effect because it compares different clinical practices under different treatments \citep{Thompson1997}. However, the average, \(E(\delta^{\crude}_p)\), over pairs is a causal effect, because the exchangeability of potential outcomes and  between clinical practices 1 and 2 (Condition 1 above) implies (proof omitted) that
\vspace*{-.5cm}

\begin{equation}
E(\delta^{\crude}_p)= E\{\mu_p(t=1)\} -E\{\mu_p(t=2)\}, \mbox{ which is } \delta^{\effect},\label{effect}
\end{equation}
Thus, one can use the estimated differences, $\hat{\delta}^{\crude}_p$, within each pair as in (\ref{firstlevel}), and expression (\ref{effect}), to estimate \(\delta^{\effect}\), either with no additional assumptions (i.e., by simply averaging $\hat{\delta}^{\crude}_p$ over pairs), or under a hierarchical second level model. \vspace{.3cm}

\noindent {\em Remark 2. }The objective meaning that the potential outcomes assign to the terms in the model (\ref{firstlevel}) implies the following, subtle fact: if the pair-specific \(\delta^{\crude}_p\) are to be eliminated (i.e., marginalized over) from the distribution (\ref{firstlevel}), then  \(\delta^{\crude}_p\) should be first integrated out of (\ref{firstlevel}) based on the conditional distribution  \(pr(\delta^{\crude}_p\mid v_p^{\crude})\), i.e.,
\begin{equation}
pr(\hat{\delta}^{\crude}_p \mid v_p^{\crude})= \int \pr ( \hat{\delta}^{\crude}_p\mid \delta^{\crude}_p, v_p^{\crude})\cdot pr(\delta^{\crude}_p\mid v_p^{\crude} )\thinspace \cdot d(\delta^{\crude}_p).\label{lkdfull}
\end{equation}
This becomes relevant when examining the existing hierarchical modeling methods.

We next discuss complications of existing methods for estimating the effect of intervention \(\delta^{\effect}\).
 We demonstrate the arguments by assessing the effect of the Guided Care intervention on the functional health outcome of the patients as measured by the physical component summary of the Short Form (SF)-36 version 2 \citep{SF36}.
 \section{Complications with existing methods}
\subsection{Consequences when ignoring covariates. }
Table \ref{table:datatable} displays the observed average SF-36 scores for each of the seven pairs of practices in the Guided Care study (see outcome rows denoted as uncalibrated).  Also displayed are the within pair differences in average SF-36 outcomes between control and intervention.

[Table \ref{table:datatable} here ]

Using these outcome data and ignoring covariates, we first obtain the estimate of the overall effect \(\delta^{\effect}\) based only on the design-derived fact (\ref{effect}) that the average of \(\delta^{\crude}_p\) equals the effect of interest \(\delta^{\effect}\) (see Table \ref{table:results}, 1st level, ``uncalibrated on covariates"). Because this first-level approach makes no further assumptions about the joint distribution of \(pr(\delta^{\crude}_p, v_p^{\crude})\), the MLE of \(\delta^{\effect}\) is simply the unweighted sample average of \(\hat{\delta}^{\crude}_p\), with its standard error estimated by the jackknife. Table \ref{table:results} also reports the permutation test of no true effect for any person, by randomly permuting the labels of treatment within each pair.

For a hierarchical second-level (meta-analytic) inference, the current approach for paired-clustered designs (e.g., \citealp{Thompson1997, Feng2001, Hill2009}) is based on integrating the likelihood in (\ref{firstlevel}) over the marginal distribution \(pr(\delta^{\crude}_p )\), to obtain:
\begin{align}
pr^{*}(\hat{\delta}^{\crude}_p \mid v_p^{\crude}, \delta^{\effect} ) &=\int pr(\hat{\delta}^{\crude}_p \mid \delta^{\crude}_p, v_p^{\crude})\cdot pr(\delta^{\crude}_p)\cdot {d} (\delta^{\crude}_p);\label{Tlkd}\\
\mbox{ where }&pr(\delta^{\crude}_p) = Normal(\delta^{\effect},v^2). \nonumber
\end{align}

Table \ref{table:results} (see 1st+2nd level, ``uncalibrated on covariates") shows inference for the effect \(\delta^{\effect}\) using the above likelihood (\ref{Tlkd}), namely, the method of  \cite{Thompson1997} with and without profiling out the variance \(v^2\) (see row 3 and 4); and also inference based on the mean of the posterior distribution of \(\delta^{\effect}\) using the uniform shrinkage prior on \(v^2\) as suggested by \cite{Daniels1999} (see row 5). For comparison, we also obtained the two-sided tail probability from the distribution of the MLE from (\ref{Tlkd}) as obtained from all the permutation possibilities of the intervention and control labels of clinical practices independently across pairs. None of these results suggest any substantial effect for the intervention.

In general, the hierarchical and non-hierarchical methods without covariates can be inaccurate for at least one of the following two reasons. First, any substantial covariate imbalances between clinical practices within a pair can result in substantial uncertainty, which is reflected in the variance of the estimators of the effect, and which may have influenced the point estimate. For the Guided Care study, Table \ref{table:covariatetable} shows that a number of covariates show substantial imbalance between intervention and control groups. For example, the continuous covariate \verb"Chronic Illness Burden" has severe imbalances between the clinical practices in pairs 2, 5 and 7, with t-statistics being $-3.07$, $-4.81$ and $2.52$, respectively.

[Tables 2, 3 here ]

The hierarchical model approach, in addition to its normality assumption, can be questioned for the following subtle reason. %{\color{red} [Why normality assumption is questionable here for Thomphson's method? We will also use this normality assumption later.]}
In order to integrate out \(\delta^{\crude}_p\) from the likelihood (\ref{firstlevel}) to obtain a likelihood that, like (\ref{Tlkd}), still depends on the variances \(v_p^{\crude}\), one must integrate \(\delta^{\crude}_p\) with respect to the {\em conditional} distribution of the estimand \(\delta^{\crude}_p\) given the variance \(v_p^{\crude}\), as in (\ref{lkdfull}) of Remark 2, and not with respect to the marginal distribution \(\pr(\delta^{\crude}_p)\) as in (\ref{Tlkd}). The comparison of (\ref{Tlkd}) to (\ref{lkdfull}) shows that (\ref{Tlkd}) implicitly assumes the following:
\vspace*{.25cm}

\noindent {\sc Condition 2. }{\em The estimand $\delta^{\crude}_p$ and the variance $v_p^{\crude}$ of  \(\hat{\delta}^{\crude}_p\) at the first level are independent across pairs \(p\).}
\vspace*{.25cm}

The motivation for using the likelihood (\ref{Tlkd}) can be traced to Thompson et al. (1997, Section 5, Paragraph 2). There, inference for the paired-clustered design is assumed to have the same random effects structure as that of  \cite{Dersimonian1986}, who also assume Condition 2 but for a design that first randomly samples subjects from the population that a pair serves and then completely randomizes them, regardless of their clinical practice. Call this simpler design, a ``paired-strata" design. We show below that violation of Condition 2 has more severe implications for the paired-clustered than for the paired-strata design.

In the paired-strata design, the observed difference, say  \(\hat{\delta}^{\prime}_p\), in average outcomes between intervention and control individuals within a pair has mean, say \(\delta^{\prime}_p\), equal to the causal effect \(\mu_p(2)-\mu_p(1)\) of (\ref{estimands}). This means that, if the intervention has no effect in any pair, i.e., the null hypotheses, $\mu_{p}(1)= \mu_{p}(2)$  for all $p$, is correct, then \(\delta^{\prime}_p\) is a constant (0) and so Condition 2 is satisfied. As a result, an approach based on (\ref{Tlkd}) is valid for testing $\mu_{p}(1)= \mu_{p}(2)$ for all $p$ because Condition 2 is correct under the null hypothesis being tested in that design.

In the paired-clustered design, however, the mean, \(\delta^{\crude}_p\), of \(\hat{\delta}^{\crude}_p\) is not a causal effect (see Remark 1 above) even if the intervention has no effect in any cluster, i.e., even if the null hypotheses, $\mu_{p,c}(1)= \mu_{p,c}(2)$ for all $p$ and $c$, is correct. In particular, under this null, the mean \(\delta^{\crude}_p\) is \(\mu_{p,1}(1)-\mu_{p,2}(1)\), i.e., the difference between clinical practices 1 and 2 if they had both been assigned control. In practice, even after matching, the two clinical practices are expected to have imbalances in characteristics of the patients or the doctors, so that \(\delta^{\crude}_p\) is expectedly not zero, and, hence, Condition 2 can be violated. We then have the following result (proof in Appendix):\longpage
\vspace*{.25cm}

\noindent {\sc Result 1. }{\em If the intervention has no effect, $\mu(1)=\mu(2)$, but Condition 2 is violated, then the MLE of the causal effect \(\delta^{\effect}\) based on (\ref{Tlkd}) can converge to a non-zero value as the number of sampled practices increases.}
\vspace*{.25cm}

Therefore, it is important to try to assess the plausibility of Condition 2. For the Guided Care study, Figure \ref{fig:dependence_check} (left) plots the estimated values of $\sqrt{v_p^{\crude}}$ against \(\delta^{\crude}_p\). Here there appear no noticeable warnings against independence. However, the covariate imbalances shown in Table  \ref{table:covariatetable} could still be contributing to inaccurate estimates through large variances as discussed earlier.

\subsection{Complications with existing covariate methods.}
Some existing proposals do incorporate covariates into the model for \(pr(\delta^{\crude}_p)\) on the RHS of likelihood (\ref{Tlkd}). However, these approaches stop short of addressing the goal of estimating effects of policy interest. In particular, such existing approaches (e.g., \cite{Thompson1997}, Sec.5.5, \cite{Feng2001}) define the treatment effect to be a contrast in the treatment coefficients of the posited model after conditioning on a particular value of the covariates and/or of random effects specific to the clusters. The first problem with such a treatment effect is that, its meaning is not objective: if, for example, the model is misspecified, then an effect set equal to a contrast of coefficients in the model does not have a well defined physical interpretation. The second problem is that, even if the model is correct, a treatment effect that is conditional on the covariates and/or the clusters is not usually equal to the overall effect.

\section{Addressing the Problems}
\subsection{Calibration of observed covariate differences between clinical practices. }
In order to use covariates to estimate the treatment effects in (\ref{estimands}), we propose to first construct calibrated pair-specific averages, for each treatment \(t=1,2\), in the sense that the distribution of the covariates reflected in the averages will be the same as the distribution of covariates combined from both clinical practices of the pair. Inference for these calibrated averages will then lead to inference for overall effects (\ref{estimands}) with the gained precision of accounting for the difference in observed covariates between the matched clinical practices.

This section uses notation for the following additional structure for pair \(p\):
\begin{itemize} \itemsep-.3in
\item[$\circ$]
 \(X_{p,c,k}\), for the measurement of a covariate vector before treatment administration, for the \(k^{th}\) sampled patient of clinical practice \(c\);\\
\item[$\circ$] \(G_{p,c}(x)\), for the joint cumulative distribution function of the covariate vector \(X_{p,c,k}\) in clinical practice \(c\), evaluated at value \(x\); and \(G_{p}(x)\) for the joint cumulative distribution function (evaluated at \(x\)) of the covariate vector of  a patient selected at random from pair \(p\) (i.e., from the two clinical practices of that pair, combined); \\
\item[$\circ$] \(F_{p,c}(y\mid x; t)\), for the cumulative distribution function of the potential outcome \(Y_{p,c,k}(t)\) in clinical practice \(c\), evaluated at value \(y\) among covariate levels \(x\), if clinical practice \(c\) is assigned treatment \(t\); and let \(\mu_{p,c}(x;t)\), for the mean of the latter distribution.
\end{itemize}

For pair \(p\), consider now the estimable quantity, labelled as \(\mucalibr_{p}(t=1)\), that is constructed by, first, stratifying the average outcome into the covariate levels of the clinical practice \(c=1\) (assigned to treatment \(1\)), namely \(\mu_{p,c=1}(x;t=1)\), and  then re-calibrating it with respect to the covariate distribution of the two clinical practices combined (and similarly for \(t=2\)):
\begin{equation}
\mucalibr_{p,c=1}:= \int_x \mu_{p,c=1}(x;t=1) dG_{p}(x), \quad \mucalibr_{p,c=2}:= \int_x \mu_{p,c=2}(x;t=2) dG_{p}(x)\label{calibr}
\end{equation}

To understand the above estimand, consider for example two clinical practices in a pair, that, although matched as closely as possible with respect to, say, the percentage of patients with a ``low" or ``high" risk covariate (\(x=\mbox{low}\) or \(\mbox{high}\)), the percentage of low risk in clinical practices 1 and 2 is \(75\%\) and \(85\%\) respectively, i.e., still differs appreciably between the clinical practices. Suppose also that clinical practice 2 serves twice as many patients as clinical practice 1. Ignoring covariates, the quantity that can be directly estimated from the data for representing the average outcome if both clinical practices are assigned treatment \(1\) is simply the average outcome within clinical practice 1, \(\mu_{p,c=1}(1)\), which can be expressed in terms of the covariate as \(0.75 \cdot \mu_{p,c=1}(x=\mbox{low};t=1)+ 0.25\cdot \mu_{p,c=1}(x=\mbox{high};t=1)\). When using covariates, the calibrated average \(\mucalibr_{p,c=1}\) is \(0.82 \cdot\mu_{p,c=1}(x=\mbox{low};t=1)+ 0.18\cdot \mu_{p,c=1}(x=\mbox{high};t=1)\), because it generalizes the covariate-specific outcome averages under treatment 1 to the covariate distribution for both clinical practices in which \(0.75 \frac{1}{3} + 0.85 \frac{2}{3}=0.82\) have low risk.

More generally, one should expect that the calibrated contrasts \(\mucalibr_{p,c=1}-\mucalibr_{p,c=2}\), though still not equal to the target  causal effect \(\mu_{p}(t=1)-\mu_{p}(t=2)\) of (\ref{null}) in each pair, should, (a) share the property with the uncalibrated estimands, i.e., that they average over pairs to the average causal effect \(\delta^{\mbox{\footnotesize{effect}}}\) of (\ref{effect}); and (b) provide a basis for more efficient estimators than the uncalibrated contrasts. This is true if the design is more carefully formalized as follows:
\vspace{.3cm}

\noindent {\sc Condition 3. }{\em The characteristics of a clinical practice, i.e., the distribution of potential outcomes under treatments 1 and 2 conditionally on covariates, the distribution of covariates, and the number of people served by clinical practice \(c\),  namely the vector of functions \(\big [ F_{p,c}(\cdot\mid\cdot,t=1), F_{p,c}(\cdot\mid\cdot,t=2), G_{p,c}(\cdot), N_{p,c} \big ]\), is exchangeable (in distribution over pairs) between clinical practices \(c=1\) and \(c=2\). }
\vspace*{.3cm}

Then we have the following:
\vspace{.3cm}

\noindent {\sc Result 2. }{\em (a) Under Condition 3, the average over pairs of the covariate-calibrations, \(\mucalibr_{p,c=1}\), i.e.,  based on the clinical practice assigned to treatment 1 in each pair (see (\ref{calibr})) equals the average of the potential outcomes if the entire population had been assigned treatment 1 (similarly for treatment 2); hence the estimable contrast
\begin{equation}
E\{\mucalibr_{p,c=1}\}\mbox{ vs. }E\{\mucalibr_{p,c=2}\}\label{estimands2}
\end{equation}
equals the causal contrast (\ref{estimands}); (b) if \(\mu_{p,c}(x; t=c)\) are known, then the MLEs of \(E\{\mucalibr_{p,c=1}\}\) in (\ref{estimands2}) (and of the target estimands $\mu(t)$ in (\ref{estimands}), due to (a) and the invariance property of the MLE) are the averages, over the observed pairs, of the empirical analogues of (\ref{calibr}):
\begin{equation}
% \hat{E}_{{\footnotesize \mbox{all \(k\) in \(p\)}}}\thinspace\{ \mu_{p,1}(X_k, 1)\}\mbox{ and } \hat{E}_{{\footnotesize \mbox{all \(k\) in \(p\)}}}\thinspace \{\mu_{p,2}(X_k, 2)\}\label{estimands2}
\int\mu_{p,c}(x; t=c)\mathrm{d}\hat{G}_p(x), c=1,2,
\end{equation}
where \(\hat{G}_p\) is the weighted empirical distribution of covariates in pair $p$ (the weight is determined by $N_{p,c}$).
}

Condition 3 implies Condition 1. The proof of Result 2 (a) follows by iterated expectations; the proof of (b) follows because the empirical distribution \(\hat{G}_p(x)\) as defined above is, under no other assumptions, the MLE of \(G_p(x)\).

\vspace*{.3cm}
%\subsection{Inference}
In practice, and simplifying the notation for the estimable averages \(\mu_{p,c}(x; t=c)\) to \(\mu_{p,c}(x)\), one can consider modelling \(\mu_{p,c}(x)\) for each (pair \(p\), cluster \(c\)), with \(\mu_{p,c}(x;\theta)\), where
\begin{equation}
h\{ \mu_{p,c}(x,\theta)\}=\theta_{p,c}+\theta_{cov}'\cdot x\quad \mbox { and }\quad h \mbox{ is a link function.}\label{model}
\end{equation}
Since these models condition on the pairs and clusters, the parameter \(\theta\) can be estimated by weighted least squares estimator \(\hat{\theta}\), based on the first moment residuals \(Y_{p,c,k}-\mu_{p,c}(X_{p,c,k},\theta)\), where approximately
\begin{equation}
\hat{\theta}\mid \theta, \Sigma_{\hat{\theta}}  \sim Normal(\theta,\Sigma_{\hat{\theta}} ),\label{param}
\end{equation}
and where \( \Sigma_{\hat{\theta}}\) is the true variance-covariance matrix of $\hat{\theta}$, which can be estimated by the robust variance-covariate matrix denoted by \(\hat{\Sigma}_{\hat{\theta}}\).

Based on these, the calibrated estimands in (\ref{calibr}) can be estimated within each pair and clinical practice, by
\begin{eqnarray}
\widehat{\mucalibr_{p,c}} & = & \int\mu_{p,c}(x,\hat{\theta})\mathrm{d}\hat{G}_p(x), \mbox{for all } p, c,
\end{eqnarray}
whose joint distribution can be approximated by the delta method as
\begin{equation}
\text{level~}1: ~~\begin{bmatrix}\widehat{\mucalibr_{p=1,c=1}}& \widehat{\mucalibr_{p=1,c=2}} \\
\vdots & \vdots \\
 \widehat{\mucalibr_{p=N,c=1}}& \widehat{\mucalibr_{p=N,c=2}} \end{bmatrix}
\mid \theta,  \Sigma_{\hat{\mu}^{\mbox{\tiny\sf calibr}}}
 \sim {Normal}\left\{
 \begin{bmatrix}\mucalibr_{p=1,c=1} & \mucalibr_{p=1,c=2}\\
 \vdots & \vdots\\
 \mucalibr_{p=N,c=1} & \mucalibr_{p=N,c=2} \end{bmatrix}, \Sigma_{\hat{\mu}^{\mbox{\tiny\sf calibr}}}
 \right\},\label{firstlevelcal}
 \end{equation}
and where \(\Sigma_{\hat{\mu}^{\mbox{\tiny\sf calibr}}}\) can be estimated by \(\hat{\Sigma}_{\hat{\mu}^{\mbox{\tiny\sf calibr}}}\). %[check adequacy of independence. Pairwise checking and need some justification.]t
%In a paired cluster randomized trial, direct estimates of original quantities of interest, for example, probability of recovery or average health evaluation score, will help investigators to directly assess clinical practiceal benefits of new intervention. On the other hand, evaluation of null hypothesis about contrast of transformed original quantities of interest, e.g. logit of proportions, can assist decision making in light of data at hand.

% not here The covariate-calibration procedure introduced above reduces bias of treatment effect estimate and, when most important covariates are adjusted, yields nearly correct Type-I error rate. We now describe the details of two paths of inference. Let $\mu(t):=E(\mu_p(t))$ for $t=1,2$.

\subsection{Estimation of quantities of original interest}
\label{sec:orig}
Expression (\ref{firstlevelcal}) can be used for estimation of the causal contrast \(\mu(1)\) vs. \(\mu(2)\) (because of Result 2(a)); here we focus on \(\delta^{\effect}=\mu(1)-\mu(2)\).  Specifically, setting ${\delta}^{\calibr}_p=\mucalibr_{p,c=1}-\mucalibr_{p,c=2}$ and  $\hat{\delta}^{\calibr}_p=\widehat{\mucalibr_{p,c=1}}-\widehat{\mucalibr_{p,c=2}}$ we can consider the first or both levels of the following two-level model
\begin{alignat}{4}
\text{level~}1':
 &~~\begin{bmatrix}
\hat{\delta}^{\calibr}_1 \\
\vdots\\
\hat{\delta}^{\calibr}_N
\end{bmatrix}\mid
\begin{bmatrix}
{\delta}^{\calibr}_1 \\
\vdots\\
{\delta}^{\calibr}_N
\end{bmatrix},\theta,  \Sigma_{\hat{\delta}^{\mbox{\tiny\sf calibr}}}
 &~~\sim~~&  {Normal}\left\{\begin{bmatrix}
{\delta}^{\calibr}_1 \\
\vdots\\
{\delta}^{\calibr}_N
\end{bmatrix}, \Sigma_{\hat{\delta}^{\mbox{\tiny\sf calibr}}}  \right\},
\label{orig1}\\
\text{level~}2':
 &~~~~~~~{\delta}^{\calibr}_p\mid \delta^{\effect},\tau^2
 &~~\sim~~&  Normal(\delta^{\effect},\tau^2), ~~p=1,\dots,N,\label{orig2}
\end{alignat}
where expression (\ref{orig1}) follows from (\ref{firstlevelcal}), and the covariance matrix \(\Sigma_{\hat{\delta}^{\mbox{\tiny\sf calibr}}}  \), obtained by the delta method, can be estimated by \(\hat{\Sigma}_{\hat{\delta}^{\mbox{\tiny\sf calibr}}}\).

Table \ref{table:datatable} shows the results for the calibrated estimates as derived from expressions (\ref{firstlevelcal}) and (\ref{orig1}) (see rows for outcome ``calibrated on covariates") for each of the seven pairs in the Guided Care study. The covariates that are involved in the calibration are listed in Table \ref{table:covariatetable}. It is notable that these calibrated differences, \(\hat{\delta}^{\calibr}_p\), are positive, in favor of the control condition, for all pairs \(p\).

Using these, Table \ref{table:results} also reports the estimate of the overall effect \(\delta^{\effect}\), first based only on the design-derived fact Result 2(a) that the average of \(\delta^{\calibr}_p\) equals the effect of interest \(\delta^{\effect}\) and on the estimation of each of \(\delta^{\calibr}_p\) by \(\hat{\delta}^{\calibr}_p\) as in (\ref{orig1}) (see 1st level, ``calibrated on covariates"). As with the uncalibrated first-level approach, this first-level calibrated approach makes no further assumptions about the joint distribution of \(pr(\delta^{\calibr}_p, \Sigma_{\hat{\delta}^{\mbox{\tiny\sf calibr}}})\), and the MLE of \(\delta^{\effect}\) is the unweighted sample average of \(\hat{\delta}^{\calibr}_p\) (here, its standard error is estimated by the jackknife, although in general it is difficult to trust a normal approximation with seven pairs). For this reason, we also calculated the significance level of the MLE by permutation of the treatment labels, thus testing the hypothesis of no true effect in any person. In this case, and because all calibrated estimated differences have the same sign, the permutation based significance level is \(2/(2^7)=0.016\) in favor of the control condition.

For a two-level approach based on (\ref{orig1}) and (\ref{orig2}), one can estimate $\delta^{\effect}$, by first obtaining the marginalized likelihood, say, $L(\delta^{\effect}, \tau^2, \Sigma_{\hat{\delta}^{\mbox{\tiny\sf calibr}}}  )$. Then we estimated $\delta^{\effect}$ by (i) the MLE after $\hat{\Sigma}_{\hat{\delta}^{\mbox{\tiny\sf calibr}}}  $ replaces $\Sigma_{\hat{\delta}^{\mbox{\tiny\sf calibr}}}  $; (ii) the MLE after profiling $\tau^2$ out; and
%[how? the $\tau^2$ is still a function of $\delta^{\effect}$]
%The estimator is
%\begin{equation}
%\hat{\delta}^{\effect}=\left(\sum_{p=1}^N\hat{\delta}^{\footnotesize \text{calibr}}_{p}Q_{\bullet,p}\right)/\sum_{j,k}Q_{jk},\label{origest}
%\end{equation}
%with variance estimate $1/\sum_{j,k}Q_{jk}$,
%where $Q_{jk}$ is the $(j,k)$th entry of $Q:=\left[\hat{\tau}^2\mathbb{I}_{N\times N}+\Sigma^{\footnotesize \text{orig}}_{\hat{\theta}}\right]^{-1}$, and $Q_{\bullet,p}$ is its $p^{\text{th}}$ column sum;
(iii) the posterior distribution of $\delta^{\effect}$ using noninformative priors for $\tau^2$ and $\delta^{\effect}$. We use a uniform shrinkage prior for the second-level variance $\tau^2$ advocated by \cite{Daniels1999}. These results for the two-level approach are given in Table  \ref{table:results} (see rows  1st+ 2nd level; MLE, pMLE, and Bayes, respectively).

As with the uncalibrated approach, the marginalized likelihood that uses (\ref{orig1}) and (\ref{orig2}) assumes that \({\delta}^{\calibr}_p\) is independent of \(\Sigma_{\hat{\delta}^{\mbox{\tiny\sf calibr}}}  \). Figure 3, right panel, plots estimates of the square root of the diagonals of \(\Sigma_{\hat{\delta}^{\mbox{\tiny\sf calibr}}}  \), \(\sqrt{v_p^{\calibr}}\), versus estimates of \(\delta^{\calibr}_p\). Although the plot can be to some degree affected by measurement error, the \(R^2\) of 0.20 suggests that some dependence exists. Although this dependence could be modeled in a modified second level, it is unclear how convincing such an approach would be as it would introduce even more modeling assumptions. To avoid this, we calculated instead the significance level of the two-level MLE estimate when evaluated from the permutation distribution of the treatment labels.
\subsection{Assessment of the hypothesis of no effect}
\label{sec:test}
The proposed approach, in addition to being robust for hypothesis testing when evaluated by permutation, is likely to have a more general robustness property analogous to the one arising in a simpler design. Specifically, in the design of complete randomization of units (unpaired, unclustered),  \cite{rosenblum2010simple} have shown that a certain class of parametric models for covariates yield MLEs for the causal effect that are consistent for the null value if indeed there is no effect on any person, even if the models are incorrect. \cite{shinohara2012broad} showed that an extended class of models has this robustness property if the models satisfy an easy to check symmetry criterion.

For the matched-paired clustered-randomization design, analogous classes of models with such robustness property may also exist. Specifically, suppose that, more generally than model (\ref{model}), we conceptualize a parametric model as one that allows distributions \(m_{p,c}(y\mid x)\) for the outcome at value \(y\) given covariate at value \(x\) for each (pair,cluster) labelled (\(p,c\)). Many flexible models \(m_{p,c}(\cdot\mid \cdot)\) (or, for brevity, \(m_{p,c}\)), including (\ref{model}), have the property that if, for two pairs and their clusters
\[ \left( \begin{array}{cc}
p_1c_1 & p_1c_2 \\
p_2c_1 & p_2c_2
\end{array} \right),
\mbox{\normalfont{the model allows the distributions}}
\left( \begin{array}{cc}
m_{1,1} & m_{1,2} \\
m_{2,1} & m_{2,2}
\end{array} \right)
\]
\noindent then it also allows the distributions
\[ \left( \begin{array}{cc}
m_{2,2} & m_{1,2} \\
m_{1,2} & m_{2,2}
\end{array} \right)
\mbox{\normalfont{and}}
\left( \begin{array}{cc}
m_{1,1} & m_{2,1} \\
m_{2,1} & m_{1,1}
\end{array} \right).
\]
The intuition of this property is that the model allows exchangeable distributions between any two observed pairs. Following a similar reasoning to that of \cite{shinohara2012broad}, we hypothesize that if (a) there is no effect of intervention in the distribution of any cluster, i.e., in the true distributions defined in Condition 3, \(F_{p,c}(\cdot\mid\cdot ; t=1)=F_{p,c}(\cdot\mid \cdot; t=2)\) for all \(p,c\), and (b) a model that has the above symmetry property is used,  then the limit of the MLE of the causal effect (\ref{estimands2}) is null even if the model is incorrect. A detailed treatment of this issue can allow for combining validity with increased efficiency in such designs as well.

\section{Discussion}
For the design that matches clusters of units and assigns interventions to clusters within pairs, we proposed an approach that estimates the average causal effect while also explicitly calibrating possibly covariate imbalance between the clusters. The approach can use only one level of inference, or can be used in a hierarchical model.

In the Guided Care study, a first-level inference with the new approach reports estimates of the causal effect with smaller estimated variance than without using covariates (see Table \ref{table:results}). Although it is difficult to know if this is objectively true in this small sample of pairs, the results from the permutation tests between the two approaches are also consistent with this conclusion. A simple two-level approach, with or without covariates, makes an implicit assumption which can invalidate causal comparison of the interventions, and explicitly addressing the assumption would introduce additional modeling. The covariate-calibrated approach reports that the control condition leads to higher, albeit clinically insignificant, average overall SF36 score compared to that under Guided Care Nurse intervention, using either a single-level (approximate or permutation-based) analysis or a two-level permutation-based analysis.

The proposed approach is expected to be more generally robust to model misspecification when assessing the hypothesis of no effect, if the model (\ref{model}) belongs in a relatively broad class. This expectation needs formal verification, but, if confirmed, can lead to more efficient estimation, and, hence, more efficient use of resources.

An alternative to the proposed approach can be to break the matching and then use regression-assisted \citep{donner07} or doubly-robust estimators \citep{rosenblum2010simple} to estimate the treatment effect. Based on Rubin's \citep{Rubin1978} theory, the matched design is still ignorable (and so the matching can be broken) if these variables that were used to create the matching are still available and are included in the outcomes model. In contrast, if these variables are not used in the model, then the design is not ignorable if the matching is broken, and this can generally lead to bias at least in the expression of the uncertainty in inference.

\section{Supplementary Materials}
The R code that implements the method in this paper is available with this paper at the \textit{Biometrics} website on Wiley Online
Library.

\section{Acknowledgements}
The authors thank the editor, an associate editor, and two reviewers for their helpful comments that improved the presentation
of the methodology, and the NIH for partial financial support.

%Let $V$ be a $n$ by $n$ matrix with diagonals $v^2+\left[\widehat{\Sigma}^{\footnotesize \text{test}}\right]_{ii}$ and off-diagonals $\left[\widehat{\Sigma}^{\footnotesize \text{test}}\right]_{ij}$. From likelihood (\ref{test1}) and (\ref{test2}), the MLE of $\eta$ is
%\begin{equation}
%\hat{\eta}=\left(\sum_{p=1}^N{\phi_{p}(\hat{\theta})}[V^{-1}]_{\bullet,p}\right)/\sum_{j,k}[V^{-1}]_{jk},
%\end{equation}
%where $[V^{-1}]_{jk}$ is the $(j,k)$th entry of matrix $V^{-1}$ and $[V^{-1}]_{\bullet,p}$ is $p^{\text{th}}$ column sum. The variance estimate of the estimator is $\widehat{\text{Var}(\hat{\eta})}=1/\sum_{j,k}[V^{-1}]_{jk}$.

\bibliographystyle{biom}
\bibliography{paircluster-refs}

\newpage
\begin{center}
{\sc Appendix}
\end{center}

\noindent{\em Proof of Result 1. }We show that the MLE of \(\delta^{\effect}\) based on the standard meta-analytic likelihood (\ref{Tlkd}) is generally inconsistent. To do this, consider the simple but informative case of a population of pairs of practices as shown in Fig. 2, where \(\mu\) follows the positive half of the standard normal distribution across such pairs. Because \(\delta^{\crude}_p\) is \(\mu\) or \(-\mu\) with probabilities \((\frac{1}{2},\frac{1}{2})\), marginally the normality of the distribution of \(\delta^{\crude}_p\) at the second level of (\ref{Tlkd}) holds with \(\delta^{\effect}(= E(\delta^{\crude}_p))=0\) and with \(\mbox{var}(\delta^{\crude}_p)=1\). Consider also, for simplicity, that  \(\mbox{var}(\delta^{\crude}_p)\) is known, and that within clinical practices, the number of patients sampled is a constant \(n\) and the variances \(\sigma^2_{p,c}(t)\) are known and are as given in Fig. 2. Then, the maximizer \(\hat{\delta}^{\effect}\) of the likelihood in (\ref{Tlkd}) is \(\sum_{p}u_p\hat{\delta}^{\crude}_p/\sum_pu_p\) where \((u_p)^{-1}= \mbox{var}(\delta^{\crude}_p)+v_p^{\crude}\), and
\begin{equation}
v_p^{\crude}=\begin{cases}
                           w_1=\frac{2}{n}, &\mbox{if the practice \(p\) is of } \mbox{type}_p=1;\\
                           w_2=\frac{1+\sigma^2}{n}, &\mbox{if the practice \(p\) is of } \mbox{type}_p=2.\\
                           \end{cases}\nonumber
\end{equation}
The probability limit of \(\hat{\delta}^{\effect}\) is \(E(u_p\hat{\delta}^{\crude}_p)/E(u_p)\), and its sign will be the sign of \(E(u_p\hat{\delta}^{\crude}_p)\). Here, although  \(E(\hat{\delta}^{\crude}_p)=0\), Condition 2 fails because the sign of $\delta^{\crude}_p$ depends on the magnitude of the variance \(v_p\). In particular, \(E(u_p\hat{\delta}^{\crude}_p)=E\{E(u_p\hat{\delta}^{\crude}_p\mid \mbox{type}_p)\}=\frac{\mu}{2}[\{\mbox{var}(\delta^{\crude}_p)+w_2\}^{-1}-\{\mbox{var}(\delta^{\crude}_p)+w_1\}^{-1}]\) which is non zero if \(\sigma^2\neq 1\). This means that even if the null hypothesis of no intervention effect on the means is correct, the standard meta-analytic approach (\ref{Tlkd}) is inappropriate if the intervention has an effect on the variance in at least one clinical practice.

[Figure 2 here ]

\newpage
\thispagestyle{empty}
\begin{table}[H]
 \centering
 \caption[pair-specific summary]{Summary of average SF36 outcomes for uncalibrated versus calibrated approaches. The first row block displays sample sizes; the second row block displays average outcomes that are uncalibrated and calibrated, respectively.

 %{\color{red}[ZW: Numbers are averaged over 5 imputed datasets. The between-imputation variances are very small.]}
 %The ``effects size" in each pair is the difference in estimated average outcomes between control and intervention (\(\hat{\delta}^{\crude}\) or  \(\hat{\delta}^{\calibr}\)) divided by the pooled standard deviation of the outcome within that pair.
\label{table:datatable}}
 \begin{tabular}{lrrrrrrrrr}\tabularnewline\hline\hline
 \multicolumn{3}{r}{}&\multicolumn{7}{c}{pair \(p\)}\tabularnewline
 \cline{4-10}
\multicolumn{2}{r}{}&\multicolumn{1}{r}{}&\multicolumn{1}{r}{1}&\multicolumn{1}{r}{2}&
\multicolumn{1}{r}{3}&\multicolumn{1}{r}{4}&\multicolumn{1}{r}{5}&\multicolumn{1}{r}{6}&
\multicolumn{1}{r}{7}\tabularnewline
\hline
{\bf sample size}&            &&          &        &        &        &         &        &         \tabularnewline
& $n_{p,c=1}$                                &&   17     & 16      &  42   &   23 &   52  &  23   &  28 \tabularnewline
 &$n_{p,c=2}$                                 &&   38     & 44      & 43    &  33  &  42   &  31   &  43\tabularnewline
 \hline
{\bf outcome}&            &&          &        &        &        &         &        &         \tabularnewline\\[.1cm]
uncalibrated&            &&          &        &        &        &         &        &         \tabularnewline
on covariates &&&          &        &        &        &         &        &         \tabularnewline\\[-1.5cm]
&$\hat{\mu}_{p,1}(1)$ && 36.4 &36.5 &39.6& 39.1& 39.7& 33.8& 39.6\tabularnewline
&$\hat{\mu}_{p,2}(2)$ && 37.3& 36.6 &39.3 &35.3 &35.2& 36.4 &40.9\tabularnewline\\[-.2cm]
&$\hat{\delta}^{\crude}_p$ && -0.8 &-0.1 & 0.3 & 3.8 & 4.5& -2.6 &-1.3\tabularnewline
%&\text{effect size}     && -0.08 &-0.01  &0.04 & 0.39 & 0.44& -0.28& -0.16\tabularnewline
&$\left(v_p^{\crude}\right)^{1/2}$   && 2.7  &2.6 & 2.0 & 2.7 & 2.1&  2.6  &2.2\tabularnewline
\tabularnewline
calibrated&            &&          &        &        &        &         &        &         \tabularnewline
on covariates &&&          &        &        &        &         &        &         \tabularnewline\\[-1.5cm]
&$^*\hat{\mu}^{\calibr}_{p,1}$       && 37.6 &38.8 &39.5 &38.0 &38.7 &35.5& 40.9\tabularnewline
&$^*\hat{\mu}^{\calibr}_{p,2}$    &&    36.7 &35.8& 39.4 &36.0 &36.4& 35.1& 40.0\tabularnewline\\[-.2cm]
&$\hat{\delta}^{\calibr}_p$          &&0.9  &3.0 & 0.1  &1.9  &2.3  &0.5 & 0.8\tabularnewline
%&\text{effect size}&&  0.10 & 0.31&  0.01 & 0.20  &0.23&  0.05 & 0.10\tabularnewline
&$\dagger \left(v_p^{\calibr}\right)^{1/2}$      && 2.1 & 2.4 & 1.5 & 2.0 & 1.7 & 2.2  & 1.7\tabularnewline
\hline
\multicolumn{10}{l}{\footnotesize *: calibration based on $n_{p,1}$ and $n_{p,2}$ observations in pair $p$}\tabularnewline
\multicolumn{10}{l}{\footnotesize \(\dagger\): \(v_p^{\calibr}\) is the \(p\)th diagonal element of  \(\hat{\Sigma}_{\hat{\delta}^{\mbox{\tiny\sf calibr}}}\) in expression (\ref{orig1})}\tabularnewline
\end{tabular}
\end{table}

\newpage
\begin{table}[H]
 \centering
 \caption[Results from different methods]{Results from MLE, profile MLE, Bayes estimates and permutation test in the Guided Care program study. The covariates used for calibration are listed in the first column of Table \ref{table:covariatetable}; the outcome is the physical component summary of the Short Form 36 (SF36).
\label{table:results}}
 \begin{tabular}{rrccccr}\tabularnewline\hline\hline
\multicolumn{2}{l}{}&\multicolumn{1}{r}{$\hat{\delta}^{\effect}$}&\multicolumn{1}{r}{95\% C.I.}&\multicolumn{1}{r}{s.e.($\hat{\delta}^{\effect}$)}&\multicolumn{1}{r}{$\widehat{\text{var}}(\delta_p^*)$}&\multicolumn{1}{c}{\begin{tabular}{@{}c@{}}$p$-value \\ (two-sided)\end{tabular}}\tabularnewline
\hline
\multicolumn{1}{l}{\bfseries uncalibrated}&\multicolumn{1}{l}{\it 1st level}&         &                       &&         & \tabularnewline
\multicolumn{1}{l}{\bfseries on covariates}& {MLE}&    0.5     &      $(-1.4, 2.5)$                 &1.0&  $-$       &   0.59  \tabularnewline
& permutation&    $-$     &         $ -$             &  $ - $ &    $- $    &   0.61  \tabularnewline
& \multicolumn{1}{l}{\it 1st+2nd level}&         &                       &&         &     \tabularnewline
& ~~MLE&$0.6$&$ (-1.2, 2.5)$&0.9&$0.7$&$ 0.50$\tabularnewline
& ~~pMLE&$0.6$&$(-1.5, 2.7)$&$-$&$0.7$&$-$\tabularnewline
& ~~Bayes&$0.6$&$ (-1.7, 3.0)$&$1.2$&$4.3$&$ 0.60 $\tabularnewline
& ~~permutation&$-$&$-$& $-$&$-$&$0.60$\tabularnewline

\hline
\multicolumn{1}{l}{\bfseries calibrated } & \multicolumn{1}{l}{\it 1st level}&         &                       &&         & \tabularnewline
\multicolumn{1}{l}{\bfseries on covariates} & {MLE}&    1.4     &      $ (0.5, 2.2)  $                  &0.4&    $-$   &   \(<\)0.01  \tabularnewline
& permutation&     $-$    &  $-$                     &   $-$  &  $-$       &     0.02 \tabularnewline
& \multicolumn{1}{l}{\it 1st+2nd level}&         &               &&         &     \tabularnewline
& ~~MLE&$1.2$&$ (-0.2, 2.6)$&0.7&$0.0$&$ 0.08$\tabularnewline
& ~~pMLE&$1.2$&$ (-0.2, 2.6)$&$-$&$0.0$&$-$\tabularnewline
& ~~Bayes&$1.3$&$ (-0.4, 2.9)$&0.9&$1.5$&$ 0.13$\tabularnewline
& ~~permutation&$-$&$ -$& $-$ &$-$&$0.02$\tabularnewline
\hline
\multicolumn{6}{l}{\footnotesize *: represents $\delta^{\crude}_p$ for the uncalibrated approach and $\delta^{\calibr}_p$ for the calibrated approach.}\tabularnewline
\end{tabular}
\end{table}

\newpage
\begin{table}[H]
 \centering
 \caption[pair-specific summary]{Checking covariate imbalances within each pair. For a continuous covariate (indicated by (a)), we calculate effect size as difference divided by pooled standard deviation. For a categorical covariate ((indicated by (b))), odds ratio is calculated comparing rates of occurence of each category between two clusters in a pair. To prevent infinite odds ratio, $0.5$ is added to all the cells when calculating sample odds ratios.
\label{table:covariatetable}}
 \begin{tabular}{rrrrrrrr}\tabularnewline\hline\hline
 \multicolumn{1}{c}{}&\multicolumn{7}{c}{pair}\tabularnewline
 \cline{2-8}
\multicolumn{1}{r}{}&\multicolumn{1}{r}{1}&\multicolumn{1}{r}{2}&
\multicolumn{1}{r}{3}&\multicolumn{1}{r}{4}&\multicolumn{1}{r}{5}&\multicolumn{1}{r}{6}&
\multicolumn{1}{r}{7}\tabularnewline
\hline
&&                  &        &        &         &        &         \tabularnewline
\verb"age at interview"$^{(a)}$ & 0.3& -0.3& 0.1& 0.6 & 0.0 & 0.1& -0.1\tabularnewline
\verb"Chronic Illness Burden"$^{(a)}$     &  0.5 &-0.6 &0.0 &0.0 &-1.1 & 0.1  &0.6\tabularnewline
\verb"SF36 Mental"$^{(a)}$ & -0.3  &0.1 &0.3 &0.2  &0.3 &-0.6 &-0.5\tabularnewline
\verb"SF36 Physical"$^{(a)}$ &  -0.1 &-0.4 &0.1 &0.5  &0.4& -0.6 &-0.3\tabularnewline
\tabularnewline
\verb"lives alone"$^{(b)}$& 1.4& 0.8& 0.7 &0.7 &1.6 &0.9 &0.5\tabularnewline[-0.5ex]
\(>\)\verb"high school education"$^{(b)}$&0.4 &0.5 &0.7& 1.4& 0.8 &0.8& 1.1\tabularnewline[-0.5ex]
\tabularnewline
\verb"Female"$^{(b)}$ &2.4& 0.6 &1.0& 0.6& 1.0 &2.5&1.1\tabularnewline[-0.5ex]
\tabularnewline
\verb"race"$^{(b)}$ &\tabularnewline
\verb"Caucasian"&0.5 & 0.2 &0.9 &0.8& 1.5 &0.5 &0.7\tabularnewline[-0.5ex]
\verb"African American"&2.2 & 0.9 &1.2 &1.2 &0.8& 1.6 &1.2\tabularnewline[-0.5ex]
         \verb"other"&2.2 &\multicolumn{1}{r}{15.0} &1.0& 1.4 &0.6& 1.3 &1.5\tabularnewline[-0.5ex]
\tabularnewline
\verb"finances at end of month"$^{(b)}$ &\tabularnewline
\verb"some money left over"& 0.0 &0.7 &1.4 &0.7 &1.5 &0.7& 0.6\tabularnewline[-0.5ex]
  \verb"just enough to make ends meet"&8.9 &1.0 &0.3 &1.3 &0.6& 1.2& 1.4\tabularnewline[-0.5ex]
    \verb"not enough to make ends meet"&\multicolumn{1}{r}{18.2} &8.4& 7.0 &1.0 &1.2 &2.0 &1.6\tabularnewline[-0.5ex]
\tabularnewline
\verb"self rated health"$^{(b)}$ &\tabularnewline
\(\geq\)\verb"very good"&0.3& 0.3& 0.8& 2.2& 0.3& 0.8& 0.6\tabularnewline[-0.5ex]
  \verb"good"&2.6& 3.4& 1.4& 0.4 &2.5 &0.8& 1.4\tabularnewline[-0.5ex]
  \verb"fair"&0.9& 0.9 &0.4 &0.3 &2.5 &4.2 &0.5\tabularnewline[-0.5ex]
  \verb"poor"&6.8& 1.5& 3.1 &4.4 &2.0 &4.2&2.1\tabularnewline
\hline
\end{tabular}\\
\end{table}
\newpage

\vspace*{-1.5cm}

\newpage

\begin{figure}[H]
\begin{center}
\includegraphics[width=0.7\textwidth]{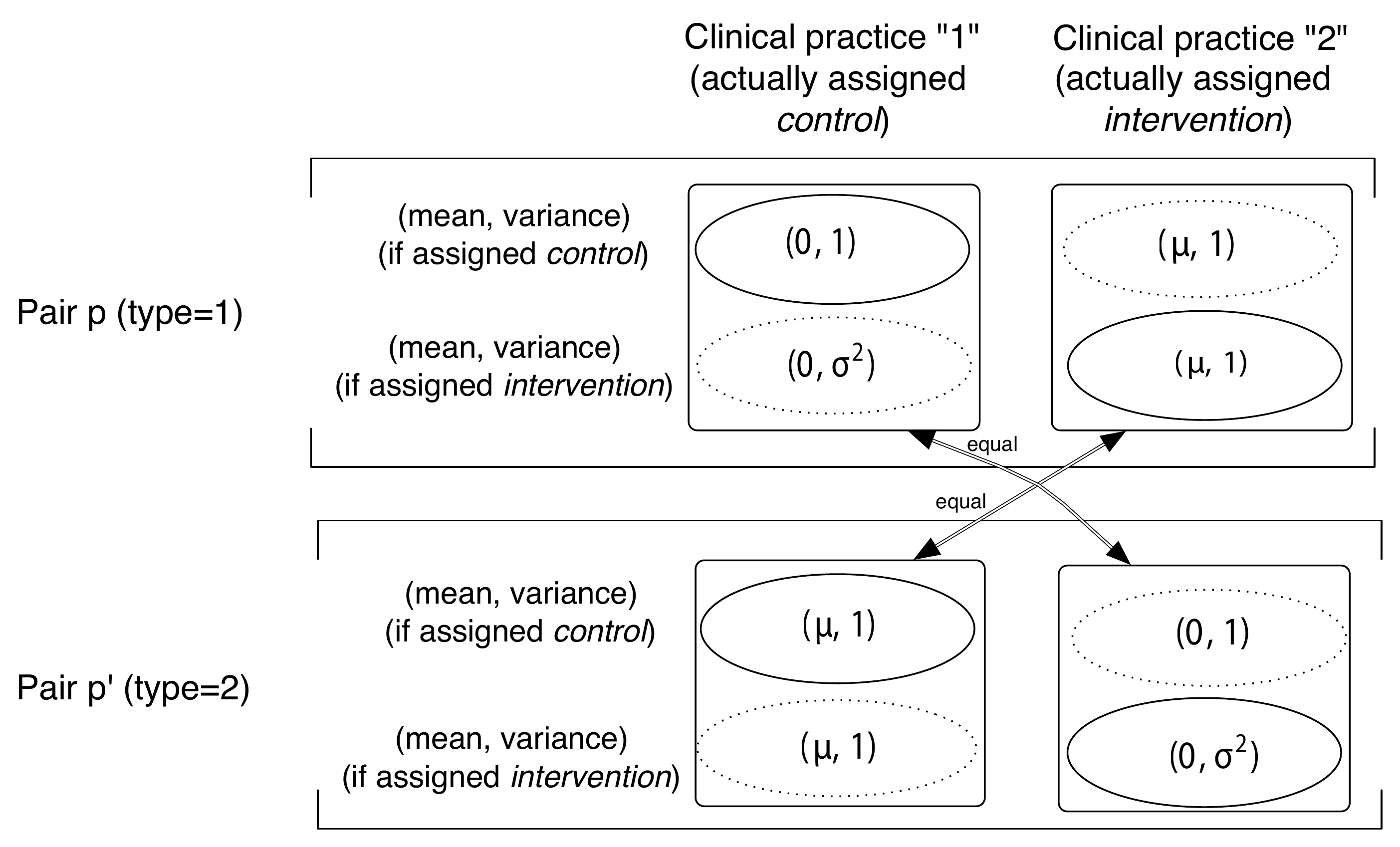}
\end{center}
\caption{Structure for the example used in the  proof of Result 1 (Appendix 1). Shown is one true type of pair and the {\em two} types of {\em observed} pairs to which it can give rise, depending on which clinical practice is assigned control. In each parentheses shown are the mean and variance of the potential outcomes of patients of the corresponding clinical practice and under a give treatment, as denoted in Fig. 1. }
\label{fig:exch}
\end{figure}

\newpage
\begin{figure}[H]
\caption{Checking second level dependence. Left: estimates of $\sqrt{v_p^{\crude}}$ versus $\delta^{\crude}_p$; Right: estimates of $\sqrt{v_p^{\calibr}}$ versus $\delta_p^{\calibr}$, where $v_p^{\calibr}$ are the diagonal elements of $\Sigma_{\hat{\delta}^{\mbox{\tiny\sf calibr}}}$.
%{\color{red} [ZW: The second graph is based on averages over 5 imputed data sets.]}
}
\label{fig:dependence_check}
\begin{center}
\hspace*{-.1cm}\includegraphics[width=14cm]{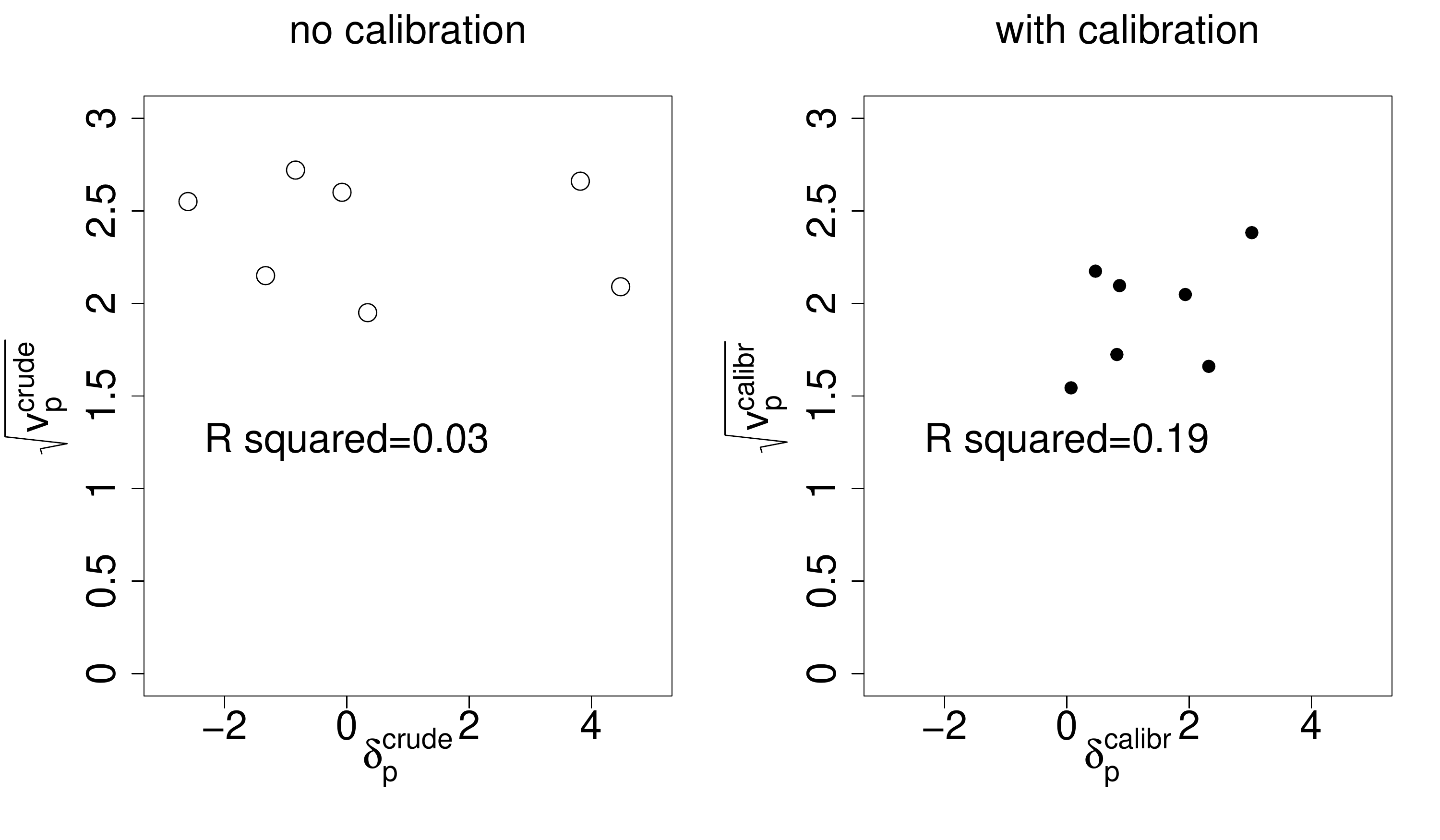}
\end{center}
\vspace{-0.1in}
\end{figure}

\end{document}